\documentclass[conference]{IEEEtran}
\usepackage{cite}
\usepackage[pdftex]{graphicx}
\usepackage{amsmath}
\usepackage{amsfonts}
\usepackage[caption=false,font=footnotesize]{subfig}
\usepackage{url}
\usepackage{booktabs}

\begin{document}
%\title{Is There a Difference in HDD Reliability by Manufacturer? Evidence From Over 400,000 HDDs}
\title{%Are There Manufacturer Differences in Hard-Drive Reliability? Evidence From Over 1.6 Million Drive-Years
Are There Manufacturer Differences in Hard-Drive Reliability?
%Manufacturer Differences in Hard-Drive Reliability
}

\author{
\IEEEauthorblockN{Christoph Siemroth}
\IEEEauthorblockA{University of Essex\\
CO4 3SQ Colchester, UK}
\and
\IEEEauthorblockN{Yeomyung Park}
\IEEEauthorblockA{Sungkyunkwan University \\
03063 Seoul, Republic of Korea} 
}

%\IEEEoverridecommandlockouts
%\IEEEpubid{\parbox{\textwidth}{\footnotesize \copyright~2026 IEEE. Personal use of this material is permitted. Permission from IEEE must be obtained for all other uses, in any current or future media, including reprinting/republishing this material for advertising or promotional purposes, creating new collective works, for resale or redistribution to servers or lists, or reuse of any copyrighted component of this work in other works.}}
\IEEEoverridecommandlockouts
\IEEEpubid{\parbox{\textwidth}{\rule{0pt}{18pt}\footnotesize \copyright~2026 IEEE. Personal use of this material is permitted. Permission from IEEE must be obtained for all other uses, in any current or future media, including reprinting/republishing this material for advertising or promotional purposes, creating new collective works, for resale or redistribution to servers or lists, or reuse of any copyrighted component of this work in other works.}}
\maketitle

\begin{abstract}
Based on the Backblaze hard disk drive (HDD) dataset, we analyze whether the four major HDD manufacturers represented in the dataset---HGST, Seagate, Toshiba, Western Digital (WD)---show differences in short- to medium-term HDD failure rates. Using two different duration regression models, we find---holding constant drive age, capacity, form-factor, and drive temperature---that Toshiba's failure rate is slightly above Seagate's. HGST HDD failure rates are the lowest, about 41\% of Seagate's. WD HDD failure rates are significantly above HGST's, but still only about 52\% of Seagate's. We also document the effects of age, capacity, temperature and drive location on failure rates.
\end{abstract}


\IEEEpeerreviewmaketitle

\section{Introduction}
The typical consumer uses only a handful of different hard disk drives (HDDs) in their lifetimes, not enough to find out by experience whether a specific HDD manufacturer is more reliable than another. Hard-drive manufacturers have a good idea about their own failure rates, given testing and returns, but they might know less about their competitors' data.   Large data storage and cloud service companies might be among the few with the necessary data to investigate manufacturer differences in reliability, since they observe the fate of thousands of HDDs. Here, we analyze the data from one of these data-centers.

The question of drive reliability is important, not just for data security but also for operating cost reasons: For data-centers and cloud storage providers with hundreds of thousands of disks, even a one percentage point drop in yearly failure rates translates into massive cost savings. This is especially true because among the major hardware parts of a server, HDDs are the most unreliable component \cite{vishwanath2010characterizing}.

In this article, we use the large Backblaze data-center dataset of daily HDD snapshots to investigate whether there are significant differences in the reliability and failure rates of HDDs between four major manufacturers, HGST, Seagate, Toshiba, Western Digital (WD).\footnote{As of 2025, three manufacturers in our sample (Seagate, Toshiba, WD) manufactured an estimated 99\% of all HDDs worldwide \cite{MRW2025,Forbes2025}. Hence, our study is very comprehensive in covering the meaningful manufacturers. HGST was a big manufacturer, but was acquired by WD in 2012 \cite{forbes2012} and later stopped making its own HDDs.} Overall, we have more than 440,000 HDDs in the sample, combined with a runtime of more than 1.66 million years, allowing us to obtain very precise estimates in the failure rates of HDDs by manufacturer, despite the low annual probability of failure of individual HDDs. Observing drives of several different manufacturers at the same company is an advantage, because the drives operate under the same conditions and procedures, thus limiting the number of confounding factors when comparing across manufacturers. Still, since drives of different manufacturers have different ages and installation years, the analysis is challenging and requires more than a comparison of annualized failure rates.

Using duration regression models, we can compare drives across manufacturers, but effectively holding drive age, capacity, form-factor, and drive temperature constant. We find that the HDDs from HGST have the lowest failure rate; about 41\% of the failure rate of Seagate HDDs. WD HDDs have about 52\% of the failure rate of Seagate HDDs, but it is significantly worse than HGST's. The Toshiba HDD failure rate is about 107\% of Seagate HDDs and hence the highest. 

Beyond statistical significance, these manufacturer differences are of high practical importance: The HGST to Toshiba differences imply more than a doubling of the failure rate, thus more than doubling the time and cost spent on replacing drives. For interpretation, the manufacturer differences we estimate are based on a dataset spanning multiple model generations, and so they capture persistent manufacturer differences beyond specific models. In addition, since only a small share of HDDs in the Backblaze dataset were installed more than 10 years ago, our estimates capture short- to medium-term failure rates.\IEEEpubidadjcol

Our goal in this article is to estimate possible manufacturer differences, not to provide a comprehensive explanation for them. However,  we can rule out  that the manufacturer differences are due to differences in drive age, capacity, form-factor, or average drive temperature, as we statistically hold these constant. We also rule out that HDD positions (as measured by HDD cluster) in the data-center affect the manufacturer differences. Moreover, since WD acquired HGST in 2012 \cite{forbes2012}, one year prior to the start of our sample, it is possible that technology and procedures were shared between the two. This would explain why their failure rates are similar, compared to the other competitors' failure rates.

Beyond manufacturer effects, we provide some evidence that drive age, HDD position/assignment in the data-center, drive temperature and capacity matter for the reliability of HDDs. Age displays manufacturer-specific patterns which are not always monotone. Larger drive temperatures increase failure rates. And higher capacity drives tend to be more reliable. We also find evidence that reliability improvements by drive generation differ by manufacturer.

Since the Backblaze data is increasingly used in the academic literature, we also document some striking features about the data. First, a large share of HDDs is removed from the dataset without failure. %Hence,  ``right-censoring" (non-observance of failure dates) occurs not only for the large share of drives that have not yet failed in the last data snapshot, but also occurs for drives that were removed from the data in the past before failing. 
Hence,  ``right-censoring"---when a drive's failure date is not observed because the drive was still functioning in the most recent data snapshot---occurs not only for the large share of drives that have not yet failed, but also occurs for drives that were removed from the data in the past before failing. 
Second, HDD manufacturers are not evenly represented across time. In 2014, HGST was the most common manufacturer among newly installed HDDs, in 2017 Seagate was the most common manufacturer among newly installed HDDs, in 2023 it was Toshiba and in 2024 it was WD. This means there are  differences in drive age by manufacturer, and serious analyses of reliability have to account for these age differences, given that drive age is a major determinant of drive failure rates. Third, the share of HDDs for which we do not observe an installation date or year---since they were in the sample from the very first day---is very small compared to the share of HDDs where the installation date can be inferred from the data. %Hence, ``left-censoring" (unobserved starting times) is not a major issue in this dataset.
Hence, ``left-censoring"---when a drive's precise installation date and hence its survival duration is not observed---is not a major issue in this dataset.

An advantage of the Backblaze data over manufacturer returns data and consumer review data is that drives of all manufacturers are treated similarly, by the same technical staff and being used for the same purposes, since they are operated by the same company. Review and returns data from different consumers and companies, on the other hand, cannot rule out that those who prefer one manufacturer use the drives very differently from those who prefer other manufacturers. Hence, there are more confounding factors with returns and review data, in addition to possible selection problems with user reviews (e.g., negative experiences may be more likely to lead to a review than positive experiences).

\section{Related Work and Contribution}
The recent literature on HDD failures has focused on predicting drive failures \cite{wang2013two}, often based on the individual drive's SMART attributes \cite{botezatu2016predicting,rincon2017disk,barelli2021unsupervised} and using machine learning methods \cite{aussel2017predictive,xiao2018disk,lu2020making,ahmed2024cost,han2025adversarial}. Hence, the focus in the literature is on reliable ahead-of-time prediction of failure during operation, rather than on quantifying the statistical factors that contribute to failures over a drive's lifetime, as we do in this study. This existing literature does not investigate manufacturer differences, and indeed aims to find prediction methods that are valid across HDD models and manufacturers. This difference in research question also motivates differences in the methodology. If useful prediction of failure is the goal, then machine learning models might be best suited. But such models are not as well suited to quantify and interpret manufacturer differences, where more classical regression techniques are appropriate. In addition to failure prediction, the literature also studies identification of slow failures, where HDDs are working but are much slower than normal \cite{lu2023perseus}. 

Outside of the academic literature, Backblaze regularly posts descriptive analyzes of HDD failure rates on their blog.\footnote{For example here for the first quarter of 2025: \url{https://www.backblaze.com/blog/backblaze-drive-stats-for-q1-2025/}.} In their posts, Backblaze  reports HDD annualized failure rates by model, though not aggregated by manufacturer. Their methods differ substantially from ours. Backblaze compares the available drives at the time. In this article, we document that certain manufacturers are overrepresented in the first cohorts of HDDs, hence the drives of these manufacturers tend to be older. Thus, simply comparing drives across manufacturers is not like-for-like. To account for this, in our regression approach, we  control for the start year of each drive, thus effectively comparing drives of similar age across manufacturers. We also control for drive form-factor, capacity, and we rule out that drive position in the data-center affects the manufacturer difference estimates. 

Moreover, unlike annualized failure rate analyses, our duration model regressions take into account information contained in censoring events (when a drive exits the data without failure), thus using the information contained in the data to the fullest.\footnote{Duration models treat durations as a random variable with a distribution $f(t)$ to be estimated. If many drives fail at time $t_1$, then $f(t_1)$ is estimated to be larger via maximum likelihood.  If a drive is censored at time $t_2$, then this contributes to the maximum likelihood function not only in that the drive has not failed for $t<t_2$---as a annualized failure rate analysis would---but it anticipates that the true duration must be $t>t_2$. Thus, this observation enters the likelihood function as $1-F(t_2)$ \cite[eq. 20.23]{wooldridge2010econometric}, where $F(.)$ is the cumulative distribution function of the duration, thus making $t>t_2$ more likely. An annualized failure rate analysis, on the other hand, does not use the information contained in censoring. Given that there are many right-censored HDDs in the sample, it is important to explicitly model and properly take these partially observed outcomes into account.} Finally, the Backblaze analyses tend to focus on the data of the most recent quarter, whereas we use the full time range from the start in 2013 for a comprehensive analysis. Thus, our contribution is a revised methodology and an analysis of the full data sample.

Several studies use data from other data-centers. Researchers  have investigated determinants of HDD failures in a large drive population at Google \cite{Pinheiro2007}, but explicitly avoid reporting manufacturer differences ``due to the proprietary nature of these data." Hence, they leave a gap that we aim to fill. They investigate drive age, temperature, and workload as determinants of drive failure. Moreover, in Meta's HDD fleet, increased age and lifetime cumulative workload both increase HDD failure rates \cite{miller2023hard}. Finally, an older study finds that, with much older drives, the  interface (SCSI, SATA, FC) makes little difference to failure rates \cite{schroeder2007understanding}.

More recently, several studies investigate factors that affect failures among SSDs, which are increasingly used at scale in data-centers \cite{Narayanan2016,xu2019lessons,maneas2020study,han2021depth}, but again without investigating manufacturer differences.

A separate strand of the literature takes a more statistical approach and investigates which distributions approximate the lifetimes and failure rates of hard-drives \cite{ye2013reliability,arslan2020distribution}.

\section{Data}

\subsection{Data Source}
We use the publicly available data from the Backblaze cloud storage provider \cite{Backblaze2025}. Their data starts in 2013, since then documenting for every day and every HDD the drive status (failure or not), several SMART attributes, as well as its unique serial number and model number (from which we infer the manufacturer). We take as given Backblaze's failure determination. %The data also includes SMART readouts for each HDD, which we do not use in this study.
In our analysis, we use all data available from Backblaze from 2013 (the earliest available) until and including the second quarter of 2025. The majority of HDDs in the dataset was added considerably after 2013, so we observe durations of more than 10 years for only 0.52\% of HDDs in the dataset. Thus, the sample size at high drive age is considerably smaller than the sample size at low ages. Consequently, our estimates reflect short- to mid-term manufacturer differences in failure rates.

\subsection{Data Preparation}
We infer the HDD installation date from the first day that a unique serial number (i.e., an individual HDD) appears in the dataset. For the drives present on the very first day in the data, we do not observe the installation date, since these drives were likely in use before the first data snapshot was made. Hence, these drives are ``left-censored", i.e., the exact starting date is not observed. For this study, we exclude the drives with unknown installation date, as it is important to control for drive age (calculated from installation date) in the regressions. Only 21,195 drives have an unknown installation date (less than 5\% of drives in the dataset).

There is a small number of SSDs in the dataset (often used as boot drives for the servers), which we exclude, as we are interested in differences of reliability within the same technology (HDDs). After excluding SSDs from the sample, there remained 4 Samsung HDDs in the dataset -- not enough for a meaningful analysis, so we excluded these as well.

\begin{figure*}[ht] %replace figure* by figure for single column figure
    \centering    \includegraphics[width=0.8\linewidth]{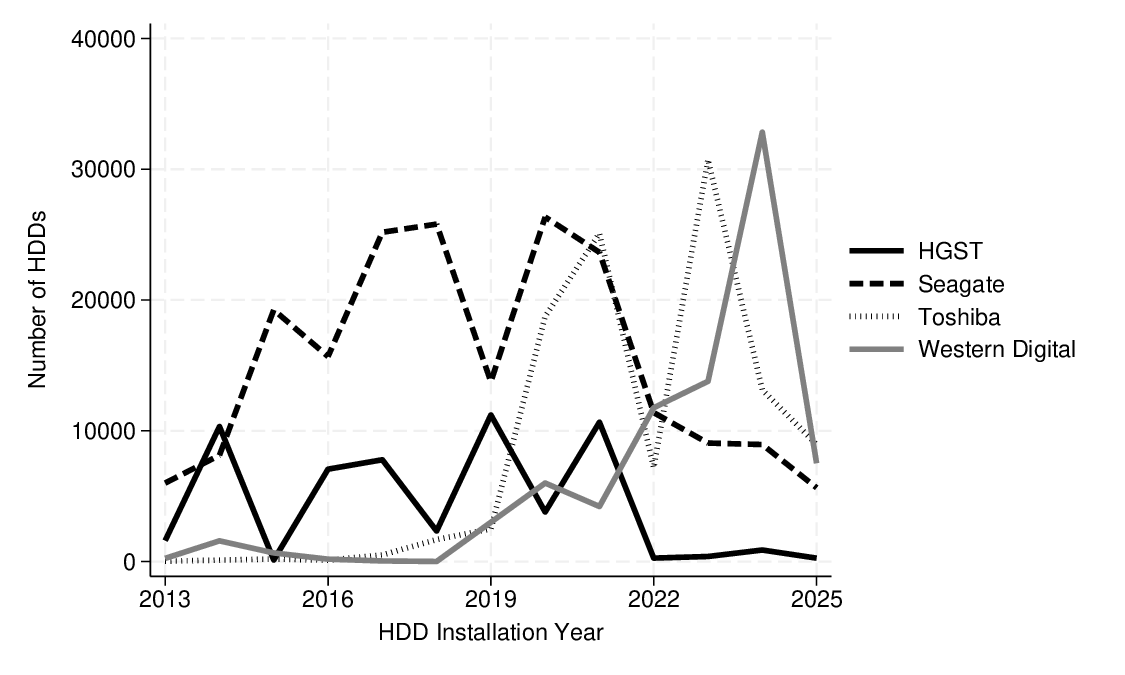}
    \caption{Number of HDDs installed by manufacturer and year}
    \label{fig:ManufacturerNumbersByYear}
\end{figure*}

The drive capacity is incorrectly recorded for a few HDDs in the data. We correct these, because we use capacity (rounded up to terrabytes) as control variable in the regressions. We correct the errors by identifying all models for which HDDs do not all have the same capacity, and then manually correct the few deviations.

A large share of drives is removed by Backblaze from the dataset without failure. In personal communication with Backblaze, we learned this is because older drives of small capacity are replaced by newer ones with larger capacity to save space and upgrade to less power consumption per capacity, which is also documented online \cite{Backblaze2024}. Overall, 146,943 drives---about 31\% of the sample---are removed without failure. We retain these in the sample, treating them as ``right-censored", i.e., not knowing their failure date, but knowing they survived at least until the removal date. This is statistically equivalent to HDDs surviving until the most recent daily snapshot in the data, where we also do not observe the (future) failure date.

Overall, we observe for every unique HDD (serial number) whether it failed or not, its survival duration in months, as well as manufacturer, capacity, form-factor (inferred from model number), installation year, and daily temperatures. The sample consists of 443,156 HDDs, of which 442,998 have a positive duration.\footnote{The difference are HDDs that were installed on the last day of the sample, hence they have not (yet) survived a single day. The duration regression models we are using only make use of observations with positive duration.} of these, We use up to 442,992 drives in the analysis, since the remaining 6 do not report temperatures.

\subsection{Manufacturer representation over time}

Figure \ref{fig:ManufacturerNumbersByYear} plots the number of installed HDDs by year and manufacturer. The figure shows that manufacturer shares are not evenly distributed across the years. Instead, HGST was the most common manufacturer for newly installed drives in 2014, whereas Seagate was the most common manufacturer in 2013 and from 2015-2020. Toshiba was the manufacturer with the most newly installed drives in 2021, 2023 and 2025. WD was responsible for the most drives installed in 2024. Overall, WD and Toshiba drives tend to be younger, as they are overrepresented in installations in more recent years, whereas HGST and Seagate drives tend to be older.

\section{Methods}

We are using duration regression models, also called survival regression models, for the analysis. As discussed in the related literature section, merely comparing failure rates between manufactures ignores that the drives have different ages (and different capacities), potentially leading to inaccurate conclusions. Hence, we use a regression approach to control for these variables, thus effectively comparing drives of different manufacturers, holding these control variables constant. Moreover, duration models are designed for this kind of application where outcomes for many units are right-censored, i.e., the full duration is not observed. In these cases, we know the drive lasted at least as long as what we observe, but not when exactly it fails, and a model for statistical inference needs to account for that. This rules out linear ordinary least squares regression for this application.

Duration models are the appropriate tool for this analysis, because they ensure a ``like-for-like" comparison in a sample where drives are not all the same age or capacity. A simple comparison of failure rates can be misleading: For instance, a manufacturer might appear more reliable simply because its drives are newer and have not yet worn out. Duration models correct for this by calculating the `hazard'---the immediate risk of failure---at every point in a drive's life. This approach allows the model to use the information from every drive in the sample, no matter how old, including those that are still functioning and where the failure date is not yet known. 

The Cox proportional hazard regression model is possibly the most popular duration model. %It is used widely in many disciplines such as economics, health, and engineering \cite{thijssens2020application}, in the latter case especially for reliability testing and estimation.
It models the hazard or failure probability within a short time period around time $t$, conditional on surviving until time $t$ and conditional on a vector of explanatory variables $\mathbf{x}$ and a vector of estimated coefficients $\boldsymbol{\beta}$, as
\begin{equation}
    \lambda(t|\mathbf{x},\boldsymbol{\beta})=\lambda_0(t) \exp(\mathbf{x}'\boldsymbol{\beta}),
\end{equation}
where $\lambda_0(t)$ is the baseline hazard function dependent on $t$ but independent of $\mathbf{x}$. The strength of the Cox model is that the function $\lambda_0(t)$ does not have to be specified for estimation of $\boldsymbol{\beta}$, hence there is no risk of misspecification of the hazard distribution. Consequently, the Cox semi-parametric model is lighter on distributional assumptions than a parametric model, its estimates are determined more by the data, and for this reason we choose it as our main model.

The coefficients of interest in this study are dummy variables denoting a specific manufacturer. %Let $\beta_1,\beta_2,\beta_3$ be dummy variables equal to 1 if and only if the HDD is made by HGST, Toshiba, WD, respectively. A Seagate drive is denoted by $\beta_1=\beta_2=\beta_3=0$. 
Let $x_1, x_2, x_3$ be dummy variables equal to 1 if and only if the HDD is made by HGST, Toshiba, or WD, respectively, and 0 otherwise. Let $\beta_1,\beta_2,\beta_3$ be the corresponding coefficients. Seagate serves as the baseline category, with its drives being represented as $x_1=x_2=x_3= 0$. 
Then the exponential of a manufacturer coefficient $\beta_i$ represents the hazard ratio (HR) of that manufacturer and of the base category Seagate, i.e., the ratio of the hazard rate of the manufacturer relative to the Seagate hazard, at the same time and holding all other explanatory variables constant:
\begin{equation}\label{eq:hr}
\begin{split}   
    \text{HR}=\frac{\lambda_0(t)\exp(\beta_1 1+\beta_2x_2+\ldots)}{\lambda_0(t)\exp(\beta_1 0+\beta_2x_2+\ldots)} =\exp(\beta_1).
    \end{split}
\end{equation}
Consequently, estimation and interpretation of the coefficients in the Cox model do not require knowledge of the baseline hazard $\lambda_0(t)$. Throughout, we estimate robust standard errors for the statistical significance tests.

In addition to the semiparametric Cox model, we estimate the flexible parametric Weibull regression model, where the hazard is assumed to be
\begin{equation}\label{weibullhazard}
    \lambda(t|\mathbf{x},\boldsymbol{\beta})=\alpha t^{\alpha-1} \exp(\mathbf{x}'\boldsymbol{\beta}),
\end{equation}
where $\alpha\in\mathbb{R},\boldsymbol{\beta}$ are estimated. If $\alpha>1$, then failure rates increase with survival time. If $\alpha<1$, the failure rates decrease with time, and $\alpha=1$ captures no duration dependence. An exponential regression model is a special case of the Weibull model with $\alpha=1$, hence we do not show exponential regression estimates separately.\footnote{As shown in Table I of the appendix, an exponential duration regression model and  Gompertz duration regression model give us almost identical estimates of manufacturer failure rate differences. Thus, several regression models with varying distributional assumptions yield almost identical results, showing robustness in terms of model choice.} %The duration dependence in our regression models is more flexible than the $\alpha t^{\alpha-1}$ term, because we control directly for the installation year of every HDD (see below).

We estimate these two different models to demonstrate that our findings do not depend on the model chosen or the functional form assumed. In each of these models, in addition to manufacturer dummy variables, we use as control variables:
\begin{itemize}
    \item Installation year: one dummy variable for every year (2013-2025), ensuring drives of the same age/installation year are compared,
    \item Capacity: rounding capacity to the next highest terrabyte, one dummy variable for every terrabyte (1TB-24TB), ensuring that drives of the same capacity are compared,
    \item Form-factor: one dummy variable for 2.5 inch drives, ensuring that drives of the same size are compared.
    \item Average HDD temperature: Based on SMART attribute 194, which measures the temperature inside the HDD case in degrees Celsius, we calculate the average daily HDD temperature. This average temperature (centered around 20\textdegree C) enters as a linear control variable, ensuring that drives with similar cooling solutions are compared.
\end{itemize}
%Moreover, we define a dummy variable Long-Term, equal to 1 for outcomes after 100 months and equal to 0 for outcomes before 100 months. By interacting all control variables with Long-Term, we estimate the effects of all variables separately for the short- to mid-term (less than 100 months) and for the long-term (more than 100 months). 

Consequently, we specify for the regressions:
% \begin{equation}
% \begin{split} 
% \mathbf{x}'\boldsymbol{\beta}=\beta_1 \text{HGST}+\beta_2 \text{Toshiba}+\beta_3 \text{Western-Digital} \\
% +\beta_4 \text{Install-2014}+\ldots +\beta_{15} \text{Install-2025} +\beta_{16} \text{Capacity-2TB} \\+\ldots +\beta_{29}\text{Capacity-24TB}   +\beta_{30}\text{Form-Factor-2.5}\\
% +\text{Long-Term}\times (\beta_{31}+\beta_{32} \text{HGST}+\beta_{33} \text{Toshiba} \\
% +\beta_{34} \text{Western-Digital}+ \ldots ).
% \end{split}
% \end{equation}
% Therefore, the baseline category is a Seagate HDD installed in 2013, with 1TB Capacity and form-factor 3.5 inch in the time window from 0 to 100 months. The differences to this base category are represented by coefficient vector $\boldsymbol{\beta}$.
\begin{equation}
\begin{split} 
\mathbf{x}'\boldsymbol{\beta}=\beta_1 \text{HGST}+\beta_2 \text{Toshiba}+\beta_3 \text{WD} +\beta_4 \text{Install-2014}\\+\ldots +\beta_{15} \text{Install-2025} +\beta_{16} \text{Capacity-2TB} \\+\ldots +\beta_{29}\text{Capacity-24TB}   +\beta_{30}\text{Form-Factor-2.5}\\+\beta_{31}\text{Average-HDD-Temperature}.
\end{split}
\end{equation}
Therefore, the baseline category is a Seagate HDD installed in 2013, with 1TB capacity, form-factor 3.5 inch and twenty degree Celsius average daily drive temperature. The hazard ratios with respect to this base category are represented by coefficient vector $\boldsymbol{\beta}$, as described in \eqref{eq:hr}.

\section{Results}

\subsection{Manufacturer differences in failure rates}

\begin{figure*} %replace figure* by figure for single column figure
    \centering    \includegraphics[width=0.8\linewidth]{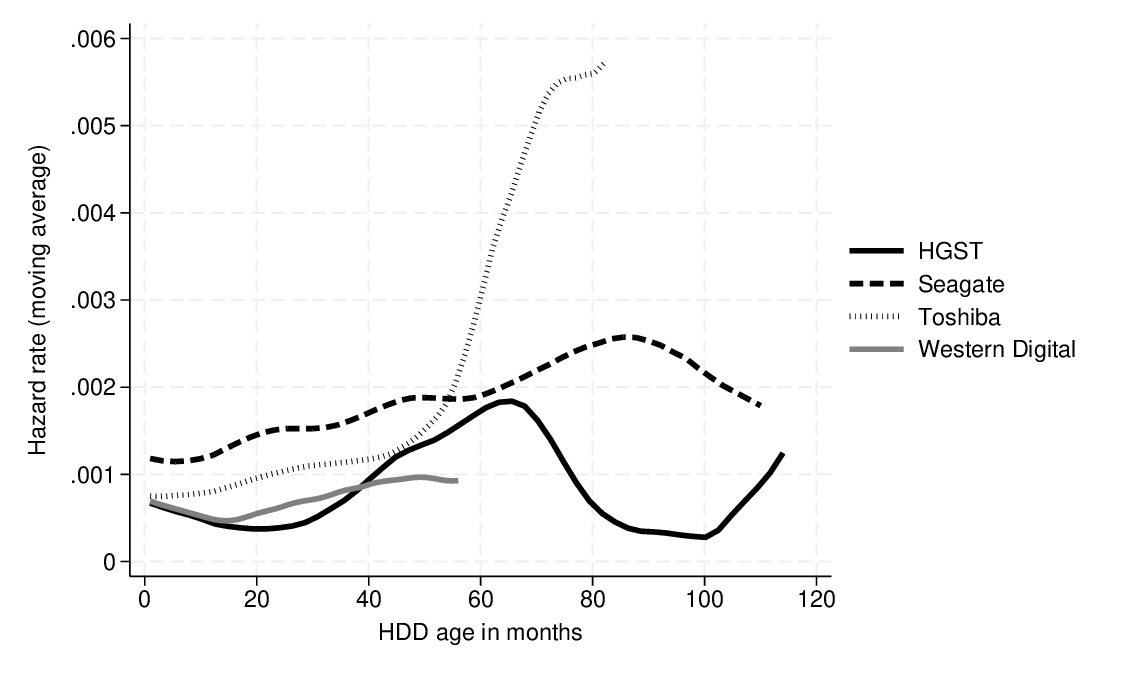}
    \caption{Hazard rate by manufacturer and drive age: Number of HDDs failed this month / number of HDDs at risk and not censored this month; discrete by month (non-parametric estimate using Epanechnikov kernel with bandwidth 5 months, making the graph a moving average). The graph only displays durations where the manufacturer has at least 1000 HDDs at risk.}
    \label{fig:HR-Truncated}
\end{figure*}

First, we plot a non-parametric estimate of the hazard (failure) rate by manufacturer and drive age in Figure \ref{fig:HR-Truncated}. Since the number of HDDs in the sample at higher ages is small, the graph is only plotted for ages where the manufacturer had at least 1,000 HDDs at risk, giving us more reliable estimates of the failure rate. As the figure shows, the length of the lines differs drastically by manufacturer: WD HDDs were added at scale only later by Backblaze, so there are fewer than 1,000 WD HDDs at age 60 months. At the same time, HGST drops down to below 1,000 HDDs at risk only at about 115 months of age.

In the range where we can compare the failure rate across all manufacturers---from months 0 to about 60---the failure rates are relatively stable across ages and the ranking between manufacturers is also quite stable. According to Figure \ref{fig:HR-Truncated}, HGST HDDs have the lowest failure rates at most ages, closely followed by WD HDDs, which do a bit better at over 40 months of age. Future data and hence longer HDDs durations will be needed to see whether WD continues to have the lowest failure rates at higher ages. Seagate HDD failure rates start out as the highest, but are overtaken by Toshiba's after about 50 months. Toshiba HDD monthly failure rates more than quadruple at age 60 months compared to early ages, a spike that no other manufacturer shows at any age in the sample. Seagate HDD failure rates are quite stable over time, whereas HGST HDD failure rates are a bit more volatile, but nevertheless largely maintain the lowest failure rates.

Regarding the level of HDD failure rates, for up to 50 months of HDD age, no manufacturer's failure rate exceeds 2 out of 1,000 HDDs within a month or (annualized) 2.4\% within a year.\footnote{The values along the lines in Figure \ref{fig:HR-Truncated} represent a weighted average of the monthly failure rate, including up to 5 months before and 5 months after, with more weight given to closer months. For details, see the caption of Figure \ref{fig:HR-Truncated}.} Between 50 and 100 months, Toshiba's failure rates jump to between 5 and 6 out of 1,000 HDDs within a month, while the failure rates of the other manufacturers tend to stay at 2.5 out of 1,000 HDDs per month or below.

Second, we get to the main statistical analysis in Table \ref{table:Main-FullSample-Temperature}. This table displays results of the two regression models described in the methods section. The numbers for HGST, Toshiba, and WD represent the estimated ratio of failure rates (i.e., hazard ratios, HR) of these manufacturers with Seagate, while holding constant the installation year (drive age), the drive capacity, the drive form factor, and average drive temperature. See equation \eqref{eq:hr} and the explanation above for the precise interpretation. Thus, Table \ref{table:Main-FullSample-Temperature} compares the failure rates across manufacturers for comparable drives.

\begin{table}[t]
\caption{HDD Manufacturer Differences in Failure Rates}
\label{table:Main-FullSample-Temperature}
\begin{center}
\begin{tabular}{lcc}
\toprule
            &     (1) Cox   & (2) Weibull   \\
 &  HR &  HR \\
\midrule
HGST        &       0.395***&       0.411***\\
            &     (0.009)   &     (0.009)   \\
Toshiba     &       1.070** &       1.073** \\
            &     (0.033)   &     (0.033)   \\
WD          &       0.514***&       0.520***\\
            &     (0.021)   &     (0.021)   \\
Average HDD Temperature&       1.021***&       1.021***\\
            &     (0.002)   &     (0.002)   \\\midrule Control Installation Year &Yes&Yes \\ Control Capacity &Yes&Yes \\ Control Form-Factor &Yes&Yes  \\

P-value of test $\text{HGST}=\text{WD}$&       0.000   &       0.000   \\
Observations&     442,992   &     442,992   \\
\bottomrule
\end{tabular}
\\ [2mm] \begin{minipage}{0.48\textwidth}
\footnotesize
{\it Note:}          Failure rate ratios/hazard ratios (HR) of various manufacturers relative to manufacturer Seagate, holding constant drive age (installation year), capacity, form factor, average drive temperature. A smaller HR is better. Robust standard errors are displayed below the HRs.          ***Significantly different from $\text{HR}=1$ at the 1\% level; **significant at the 5\% level; *significant at the 10\% level.                                             
\end{minipage}
\end{center}
\end{table}

The estimates of both the Cox regression model and the Weibull regression model are on the same scale (i.e., both are hazard ratios in an exponential specification), and the estimated values are very similar (i.e., within 2 percentage points), demonstrating robustness of the results with respect to the regression model chosen. Hence, in the following we won't discuss the estimates of the two models in Table \ref{table:Main-FullSample-Temperature} separately.

Table \ref{table:Main-FullSample-Temperature} shows that HGST has the lowest failure rate at about 39\%-41\% of Seagate's failure rate. This difference in failure rates is statistically significant at any conventional confidence level ($p<.001$). Thus, HGST has the lowest HDD failure rates among the four manufacturers. Next, WD has the second lowest failure rate, about 51\%-52\% of Seagate's failure rate. Again, this difference to Seagate is highly significant. Moreover, the significance test at the bottom of Table \ref{table:Main-FullSample-Temperature} shows that the hazard ratios (and hence failure rates) for HGST and WD are significantly different ($p<.001$). Hence, HGST's failure rate is significantly lower than WD's.

Toshiba's hazard ratio with Seagate is about 1.07, so its failure rates are estimated to be about 7\% larger than Seagate's. The difference between these two manufacturers is significantly different at the 5\% confidence level, not as clear-cut as the differences between other manufacturer pairs.

Hence, in this large sample of HDDs, with ages mostly under 10 years, we see a clear ranking: HGST has the lowest failure rate, followed by WD, both of which have about half or less of Seagate's failure rate. After this, it is close between Seagate and Toshiba, with Toshiba having slightly higher failure rates. The maximum difference between HGST and Toshiba implies more than a doubling of the drive failure rate. This suggests that the time and money spent on replacing faulty drives is more than doubled when comparing these two manufacturers. The results from the regressions in Table \ref{table:Main-FullSample-Temperature} are in line with the non-parametric values in Figure \ref{fig:HR-Truncated}.

\subsection{The effect of drive temperatures on failure rates}

According to Table \ref{table:Main-FullSample-Temperature}, average drive temperatures have a hazard ratio of 1.021; a statistically significant effect on drive failure rates. The estimated hazard ratio of 1.021 indicates that an additional degree in the average daily temperature increases failure rates by 2.1\% (not percentage points). Due to compounding, an additional 10 degrees of mean temperature is predicted to increase the failure rate by $1.021^{10}=1.231\approx 23.1\%$.\footnote{The inner 90\% of the average temperatures in our data range from  22.5\textdegree C to just above 40\textdegree C, so a 10\textdegree C change is comfortably within the observed sample temperature ranges.} Thus, suppose a drive at 40\textdegree C average temperature has a 2\% failure rate. Then according to our estimates, cooling the drive by 10\textdegree C is predicted to reduce the failure rate to $2\%\times1/1.231\approx 1.625\%$.

\subsection{Does accounting for position in the data-center change the results?}

Clearly, conditions such as the quality of the maintenance staff, or the load on the drives might affect HDD failure rates. All of these factors depend on the position and assignment of a drive in the data-center: A server dedicated to a heavy user might see  higher failure rates than another server for an inactive client. Backblaze has added position variables to their drive data only from quarter 2 and 3 of 2023. Beforehand, it is unknown where a drive was installed.\footnote{In order to include as many drives as possible in this analysis, we use the later recorded position variable values even if a drive was installed at an earlier time when these variables where not yet reported. Thus, in this analysis we only lose drives that dropped out of the sample before the position variable was reported in Q2/3 of 2023.}

\begin{table}[t]
\caption{Comparing HDD Manufacturer Differences With and Without Drive Position Controls  (Smaller Sample)}
\label{table:Robustness-ClusterControl-Temperature}
\begin{center}
\begin{tabular}{lcc}
\toprule
            &     (1) Cox   &  (2) Cox HR   \\
 &  HR &  Without  \\
\midrule
HGST        &       0.814***&       0.833***\\
            &     (0.033)   &     (0.033)   \\
Toshiba     &       0.977   &       1.029   \\
            &     (0.048)   &     (0.049)   \\
WD          &       0.442***&       0.471***\\
            &     (0.029)   &     (0.030)   \\
Cluster 20  &       1.287***&               \\
            &     (0.114)   &               \\
Cluster 31  &       1.229***&               \\
            &     (0.064)   &               \\
Cluster 40  &       0.736***&               \\
            &     (0.046)   &               \\
Cluster 50  &       1.154** &               \\
            &     (0.073)   &               \\
Cluster 52  &       0.511** &               \\
            &     (0.144)   &               \\
Cluster 60  &       0.489*  &               \\
            &     (0.203)   &               \\\midrule Control Installation Year &Yes&Yes \\ Control Capacity &Yes&Yes \\ Control Form-Factor &Yes&Yes \\ Control Avg HDD Temperature &Yes&Yes \\ 

P-value of test $\text{HGST}=\text{WD}$&       0.000   &       0.000   \\
Observations&     355,429   &     355,429   \\
\bottomrule
\end{tabular}
\\ [2mm] \begin{minipage}{0.48\textwidth}
\footnotesize
{\it Note:}          Failure rate ratios/hazard ratios (HR) of various manufacturers relative to manufacturer Seagate, holding constant drive age (installation year), capacity, form factor, average drive temperature, and drive position/cluster (only in column 1, cluster 0 is the omitted category). A smaller HR is better. Robust standard errors are displayed below the HRs.          ***Significantly different from $\text{HR}=1$ at the 1\% level; **significant at the 5\% level; *significant at the 10\% level.                                             
\end{minipage}
\end{center}
\end{table}

In this section, we re-estimate the Cox model manufacturer differences while accounting for the cluster that an HDD is part of. In Backblaze's terminology \cite{Backblaze2024}, a cluster is a collection of vaults sharing common resources (such as networking equipment), and a vault is a collection of 20 servers. Each server uses dozens of HDDs. Overall, there are 7 different clusters in the dataset. If some clusters see heavier loads, have higher temperatures or less capable maintenance staff, then by controlling for clusters in the regression, we can remove and disentangle these effects from manufacturer differences.\footnote{This is important if some clusters have proportionally more drives of a certain manufacturer: If a cluster sees heavier loads and has (say) more Seagate drives, then it would disadvantage Seagate drives.} The downside of controlling for drive position is that our sample size is reduced by about 90,000 drives, because the cluster variable is not available for drives that dropped out of the sample before this variable was reported.

Table \ref{table:Robustness-ClusterControl-Temperature} shows the estimates. In column 1, the Cox regression includes dummy variables for each of the clusters as control variables. The Cox regression in column 2 uses the same observations as column 1, but omits the cluster variable controls. Comparing the manufacturer hazard ratio estimates from column 1 and 2 in Table \ref{table:Robustness-ClusterControl-Temperature}, there is little difference, and certainly nothing that changes the ranking among manufacturers. Hence, in our case, HDD position/assignment to cluster does not change the manufacturer HDD reliability estimates a lot. However, the estimates of manufacturer differences in Table \ref{table:Robustness-ClusterControl-Temperature} (column 2) are somewhat different from Table \ref{table:Main-FullSample-Temperature}, which is solely due to the change in sample: Table \ref{table:Robustness-ClusterControl-Temperature} uses about 355,000 HDDs, whereas Table \ref{table:Main-FullSample-Temperature} uses the full sample of 442,000 HDDs. Consequently, the conclusions from Table \ref{table:Main-FullSample-Temperature} are the more credible ones: While not controlling for HDD position, we have shown this variable does not change the estimates noticeably, and Table \ref{table:Main-FullSample-Temperature} uses the larger sample.

\subsection{Drive failure rates by age, capacity, and position}

Beyond manufacturer effects, our dataset allows to make some inferences regarding other factors' influence on HDD reliability and failure rates. As seen in Figure \ref{fig:HR-Truncated}, failure rates clearly can change by drive age, and in addition, these age-specific patterns appear to differ by manufacturer. While Toshiba HDD failure rates change a lot by drive age, Seagate failure rates are more constant. Taking Toshiba as an example, failure rates can easily change by a factor of 5 over the course of the HDD lifetime. Hence, age is an important determinant of HDD failure rates, though it is not a simple linear or monotone effect, especially when considering HGST in Figure \ref{fig:HR-Truncated}.

In addition, the effect of HDD position in the data-center matters. While accounting for HDD position does not change estimates of HDD manufacturer differences (see Table \ref{table:Robustness-ClusterControl-Temperature}), there are still large differences between clusters in terms of hazard rates. Specifically, the cluster-specific estimates of the regression in column 1, Table \ref{table:Robustness-ClusterControl-Temperature} suggest that cluster 20 with the highest failure rates almost has $1.287/0.489\approx2.63$ times the failure rates of cluster 60 with the lowest failure rates -- for HDDs with the same manufacturer, capacity, age, form-factor, and temperature. This difference is even larger if we do not hold temperature constant (not shown in the table). So loads and other location-specific factors matter a lot in practice.

\begin{table}[t]
\caption{The Effect of Capacity on Failure Rates}
\label{table:Extension-Capacity-Temperature}
\begin{center}
\begin{tabular}{lcc}
\toprule
            &     (1) Cox   & (2) Weibull   \\
 &  HR &  HR  \\
\midrule
HGST        &       0.409***&       0.418***\\
            &     (0.009)   &     (0.009)   \\
Toshiba     &       0.863***&       0.875***\\
            &     (0.018)   &     (0.018)   \\
WD          &       0.595***&       0.602***\\
            &     (0.021)   &     (0.021)   \\
Capacity in TB (linear)&       0.966***&       0.961***\\
            &     (0.005)   &     (0.005)   \\\midrule Control Installation Year &Yes&Yes \\ Control Form-Factor &Yes&Yes \\ Control Avg HDD Temperature &Yes&Yes \\

P-value of test $\text{HGST}=\text{WD}$&       0.000   &       0.000   \\
Observations&     442,992   &     442,992   \\
\bottomrule
\end{tabular}
\\ [2mm] \begin{minipage}{0.48\textwidth}
\footnotesize
{\it Note:}          Failure rate ratios/hazard ratios (HR) of various manufacturers relative to manufacturer Seagate, holding constant drive age (installation year), capacity (with a linear term), form factor, average drive temperature. A smaller HR is better. Robust standard errors are displayed below the HRs.          ***Significantly different from $\text{HR}=1$ at the 1\% level; **significant at the 5\% level; *significant at the 10\% level.                                             
\end{minipage}
\end{center}
\end{table}

Drive capacity not only represents more storage volume, but also technical progress and updated manufacturing processes, since higher capacity drives are newer. To see whether higher capacity drives are more reliable, we re-estimate the Cox model from Table \ref{table:Main-FullSample-Temperature}, but include capacity as a linear term rather than a series of dummy variables as that table does. The estimates are displayed in Table \ref{table:Extension-Capacity-Temperature}. %\footnote{The manufacturer differences estimated in this table are very similar to those in Table \ref{table:Main-FullSample-Temperature}, though the latter's estimates are more reliable, as it allows for non-monotone effects of capacity. Table \ref{table:Extension-Capacity}, on the other hand, imposes a linearity/monotonicity assumption.} 
As before, this regression also holds age, manufacturer, and form-factor constant. The hazard ratio of capacity in terrabytes is estimated to be 0.966, and the difference to $\text{HR}= 1$ is highly statistically significant ($p<.001$). Hence, in this linear approximation, a drive with 1TB higher capacity is predicted to have a 3.4\% (not percentage points\footnote{Hence, suppose a 1TB drive has an annual failure rate of 2\%. Then a 2TB drive would be predicted to have a $2\%\times 0.966$ annual failure rate.}) lower failure rate. So, based on this data, modern drives are not only becoming larger in terms of storage volumes, but also more reliable.

There are at least three classes of possible explanations for this estimated improvement in reliability for higher capacity (i.e., newer) drives. First, improvements in the manufacturing process itself. Manufacturing processes improve over time as manufacturers gain experience and iterate. Thus, modern high-capacity drives benefit from improved servo technology, precision head fabrication techniques (e.g., WD switched from a dry hole process to a Damascene process, arguing it improved quality \cite{WD2017}), and quality control methodologies. Second, improvements in the HDD technology itself are likely a cause for the improvement. For example, during the sample window enterprise drives increasingly switched from air to helium-filled enclosures. Due to its lower density, helium reduces aerodynamic drag on the HDD platters, thereby decreasing mechanical stress, heat generation, and spindle bearing wear compared to conventional air-filled drives \cite{yang2010heat,aoyagi2022helium}. Other changes in HDD technology helped increase capacity, but are less likely to explain increased reliability (e.g., shingled magnetic recording uses overlapping data tracks which increase  complexity). Third, a theoretical possibility is that the requirements of HDD customers (increasingly data-centers, as workstations and retail consumers switch to SSDs) have shifted towards more reliability and a higher willingness to pay for it over time. Given the changed demand, the manufacturers may have switched to higher quality parts and larger investments in manufacturing.

Finally, whether the HDD form-factor---3.5 inch or 2.5 inch---matters for failure rates cannot be conclusively answered with the Backblaze data, because only about 0.6\% of the HDDs have a small form factor. It would require a different kind of setting and dataset to answer this question.

\subsection{Failure rates across drive generations and manufacturers}

In this section, we ask whether the gradient at which drives become more reliable over time/over the generations is different by manufacturer. Unlike in the previous section, we focus here not on capacity as a proxy for drive generation, but directly focus on installation year. While it is possible that older generation drives are installed in a given year, on the whole we expect drives to get more modern with advanced installation years. In these regressions, we will not control separately for drive capacity, as drive capacity and drive generation are highly correlated. Put differently, we want to measure the effect of newer drives on reliability, but if we were to hold capacity constant, we would not pick up the full effect.

Thus, in Table \ref{table:DifferentialTimeTrends-FullSample-Temperature}, we estimate regressions as in Table \ref{table:Main-FullSample-Temperature}, except that (1) we do not control for drive capacity, (2) we let the effect of installation year be linear rather than nonparametric via dummy variables, and (3) we allow the effect of installation year be different by manufacturer, which we achieve by interacting the installation year linear variable with the manufacturer dummy variables.

\begin{table}[t]
\caption{Differences in Manufacturer Generational Trends}
\label{table:DifferentialTimeTrends-FullSample-Temperature}
\begin{center}
\begin{tabular}{lcc}
\toprule
            &     (1) Cox   & (2) Weibull   \\
 &  HR &  HR \\
\midrule
HGST        &       0.091***&       0.097***\\
            &     (0.005)   &     (0.005)   \\
Toshiba     &       1.450***&       1.643***\\
            &     (0.117)   &     (0.127)   \\
WD          &       1.076   &       1.033   \\
            &     (0.085)   &     (0.076)   \\
Installation Year&       0.905***&       0.884***\\
            &     (0.004)   &     (0.003)   \\
HGST$\times$Installation Year&       1.420***&       1.412***\\
            &     (0.012)   &     (0.012)   \\
Toshiba$\times$Installation Year&       0.944***&       0.927***\\
            &     (0.010)   &     (0.010)   \\
WD$\times$Installation Year&       0.917***&       0.916***\\
            &     (0.010)   &     (0.009)   \\\midrule  Control Form-Factor &Yes&Yes  \\ Control Avg HDD Temperature &Yes&Yes  \\

P-value $\text{Toshiba trend}=\text{WD trend}$&       0.039   &       0.359   \\
Observations&     442,992   &     442,992   \\
\bottomrule
\end{tabular}
\\ [2mm] \begin{minipage}{0.48\textwidth}
\footnotesize
{\it Note:}          Failure rate ratios/hazard ratios (HR) holding constant form factor. Installation Year is centered on 2013, the first sample year. A smaller HR is better. Robust standard errors are displayed below the HRs.          ***Significantly different from $\text{HR}=1$ at the 1\% level; **significant at the 5\% level; *significant at the 10\% level.                                             
\end{minipage}
\end{center}
\end{table}

For the purposes of this analysis, we re-centered the Installation Year variable to 2013, the first year of the sample. To interpret the coefficients in Table \ref{table:DifferentialTimeTrends-FullSample-Temperature}, note that the manufacturer dummy variables are the hazard ratios of that manufacturer relative to Seagate \textit{at installation year 2013}, and relative to the assumed linear time trend. As we are interested in changes over time rather than manufacturer differences in a specific year, the coefficients of the manufacturer dummies are not of much interest in this table. The estimated coefficient on Installation Year is 0.905 or 0.884, depending on the regression model. Thus, Seagate drives get more reliable with each installation year, approximately by 10\%-11\%, which is quite large in magnitude. This installation year effect does not capture the same thing as the capacity effect of the previous section (which instead measured the reliability effect of an additional 1TB of capacity of a drive installed in the same year), so these estimates are not in conflict. 

Next, we can consider the interaction terms, which capture a difference of the time gradient relative to Seagate. Both the Toshiba and WD interaction terms are significantly less than 1, indicating that the improvement over time (i.e., by installation year) is better than for Seagate. To obtain the gradient for WD, we need to multiply the interaction term coefficient with the Installation Year coefficient. Thus, WD drive reliability improves by an estimated $1-0.905\times0.917=0.17$, i.e., about 17\% (not percentage points) per installation year, in this linear approximation. And it being a linear approximation, we should be hesitant extrapolating the effect beyond the sample period. Toshiba's improvement by installation year are very similar to WD's (not significantly different at the 5\% significance level in column 2), as its similar estimated interaction coefficients and the second-last line in the table indicate. Hence, Toshiba, too, sees large improvements in reliability by drive generation. Interestingly, HGST's gradient over time is worse than Seagate's, as the estimated coefficient for HGST's interaction term is significantly larger than 1. We need to keep in mind, however, that HGST was mostly installed in the first few years of the sample and later stopped making new HDDs, so any later installed HDDs are effectively older generation ones. So the shorter sample for HGST combined with a possible bias---where later HGST drive installations had to be older generation drives---explains the worse improvement gradient over time. Consequently, we do not consider the comparison between HGST and the others ``like for like" when it comes to improvements over time, though there is no issue with comparing reliability holding installation year constant, as we did in previous sections.

\section{When does it pay to buy the more reliable HDD?}
In this section, we briefly derive a relationship between the acceptable purchase price premium $P_{\text{premium}}=P_{\text{reliable}}-P_{\text{base}}$ of a more reliable drive over a base drive and the reduction in the annualized failure rate of the drive, $\Delta=\text{AFR}_{\text{base}}-\text{AFR}_{\text{reliable}}$. This relationship can then be used to answer whether buying the more reliable drive is justified.

We start with the Total Cost $TC$ of a drive over its lifetime $L$ (in years), which is the sum of its purchase price $P$ and the expected cost due to failures over the lifetime:
$$TC=P+\text{AFR}\times C\times L,$$
where AFR is the annual failure rate and $C$ is the cost of drive failure\footnote{Assuming warranty, this is the labor and logistics cost of replacement plus expected costs due to downtime.}.

From a business perspective, reliability has the benefit of lower expected replacement costs and the possible downside of a higher purchase price. The total cost of the reliable drive is less than the total cost of the base drive if and only if $TC_{\text{reliable}}\le TC_{\text{base}}$, and rearranging for the price premium:
\begin{equation}\label{maxprice}
    P_{\text{premium}}\le \Delta\times C\times L.
\end{equation}
Clearly, with longer lifetimes $L$, reliability differences have a longer time to pay off, justifying larger initial price premia. And with lower replacement costs $C$, only smaller initial price premia are justified. Thus, inequality \eqref{maxprice} gives us a simple tool to trade off procurement costs and better reliability.\footnote{Clearly, the model behind \eqref{maxprice} could still be further extended by allowing replacement costs to differ by manufacturer (relevant if not the entire lifetime is covered by warranty) or by introducing discounting of future costs.}

To use this expression in an example, suppose the lifetime of a drive (after which it will be replaced) is $L=10$ years and the cost of replacement is $C=\$100$, then the AFR difference from Seagate (around 2\%, Figure \ref{fig:HR-Truncated}) to HGST is $\Delta=0.02(1-0.41)=0.0118$, using our regression estimate of HGST having a failure rate that is 41\% of Seagate's. The maximum acceptable price premium for an HGST drive over Seagate then would be $\$100\times 0.0118\times 10 = \$11.8$. And while this number may seem small, suppose the drives actually have the same price, so choosing the less reliable one for the drive fleet costs over \$3m more in terms of total cost in the case of Backblaze with 300k active drives just due to more replacements. 

\section{Conclusion}
We provide estimates of manufacturer differences in HDD reliability. When comparing drives of the same age, capacity, and form-factor, HGST HDDs are the most reliable, whereas Toshiba is the least reliable, closely followed by Seagate. The HGST failure rate is less than half of Seagate's, whereas WD's failure rate is about half of Seagate's. We can rule out that these manufacturer differences are due to different positions (as measured by cluster) in the data-center, due to differing temperatures, due to different ages, different capacities, or different form-factors, as we account for these variables in the statistical analysis. 

Our findings should be interpreted and used with care. First, we have no data that consistently tracks drive read/write workloads across drives of different manufacturers, because different manufacturers use different SMART attributes and implement them differently, thus making workload comparisons across manufacturers difficult. Hence, we cannot directly confirm that drives of all manufacturers of the same age have similar workloads. Our cluster-location analysis indicates that accounting for drive position (which also determines tasks and workload) does not meaningfully change our manufacturer differences estimates, which is reassuring. But absent more direct workload data, our analysis is based on the informed assumption that workloads do not drastically differ between manufacturers.
Second, Backblaze, like other data-centers, largely uses enterprise drives. Hence, our conclusions apply to enterprise drives, and manufacturer differences might be different among consumer drives. Additionally, the dataset contains HDD models from over 10 years ago, and manufacturer differences among the newest models (which did not yet have time to prove themselves) might be different. That said, we see large manufacturer differences in a sample spanning several model generations, so the manufacturer differences we obtained cannot be explained by single models or single model generations. %At a time of quickly increasing technician salaries, our results suggest there may be ways to control data-center costs by substituting from HDDs with higher failure rates---requiring more staff time---to HDDs with lower failure rates.

%A large dataset with consumer drives would be needed to investigate this question. In addition, it would be interesting to conduct similar studies with SSDs rather than HDDs, though data-centers still largely use HDDs due to their larger capacities. Hence, SSD data with large populations from the same user might be hard to obtain. Finally, a more direct comparison of SSDs and HDDs in terms of reliability would be interesting, but such a comparison would have to ensure that both kinds of drives are used in a similar way, which is often not the case in practice (e.g., Backblaze uses SSDs as server bootdrives for rapid access, whereas HDDs are used for storage of the customer data).

Our results are of interest to data-centers and cloud storage providers, where even small differences in failure rates translate into large costs due to their scale. In Backblaze's case of over 300,000 active HDDs (which is still small compared to Amazon, Microsoft, Google, etc.), a doubling of the annual failure rate from about 1\% (Toshiba, Figure \ref{fig:HR-Truncated}) to 2\% (Seagate) requires replacing an additional 3,000 HDDs  per year. These failed HDDs may be under warranty, depending on the contract, so no additional hardware cost might apply. Yet the staff time required for replacement is not covered by warranty. Consequently, if salaries for data-center technicians increase faster than price differences in HDD models, then it may be increasingly beneficial to shift to more expensive but more durable HDDs.

\section*{Acknowledgment}

We would like to thank Stephanie Doyle from Backblaze, who patiently answered our questions about the data and operations at Backblaze. We thank Backblaze for publicly sharing the data.
We are grateful to Youngik Eom, Hyungmin Cho, and Panagiotis Kanellopoulos for their helpful comments and advice.



%\bibliographystyle{IEEEtran}

\bibliographystyle{IEEEtran}
\bibliography{literature.bib}

\end{document}